# Truly SubNyquist Multicomponent Linear FM Signal Decomposition Method

by Huiguang Zhang

*Email:* zhanghuiguang@stu.haut.edu.cn

December 5, 2024

## Abstract

Accurate extraction of multicomponent linear frequency modulation (LFM) signal parameters, such as onset frequency, linear modulation frequency, amplitude, and initial phase, is of great importance in the fields of ISAR, cognitive radio, electronic countermeasures, and star-ground communications. However, the task of accurately extracting the characteristic parameters of a signal is challenging when it has an extraordinarily large bandwidth as well as cross or neighboring components in the time-frequency domain. In this paper, we first review the main current methods used for multicomponent LFM signal decomposition and their challenges, and then propose a novel multi-parameter feature parameter extraction algorithm. The algorithm realizes the direct and accurate extraction of the feature parameters of multicomponent LFM signals at ultra-low sub-Nyquist sampling rate for the first time. Moreover, the algorithm is optimized for the computational complexity and anti-noise problems in practical applications, so that it has high accuracy, high efficiency and good noise robustness. We also compare the algorithm with innovative and existing methods, and the results show that the algorithm has excellent performance in feature parameter extraction accuracy, noise immunity and computational speed.

**Keywords:** Multi-component linear FM, Multi-parameter eigenparameters, Truly Sub-Nyquist

# 1 Introduction

## 1.1 Current major methods for Multicomponent LFM signal decomposition and their challenges

Linear FM (LFM) signals are the most important waveforms in low-intercept-probability (LPI) radars[1], synthetic aperture radars[2], and modern communication systems[3]. Multi-component LFM signals, on the other hand, are synthesized signals consisting of several linear FM signals with different starting frequencies and FM parameters[4][5]. This type of signal has the advantages of high range and resolution for applications in the field of ranging, and large bandwidth, high immunity to interference, and difficulty in interception for applications in the field of communications[6]. Therefore accurate estimation of parameters of multicomponent LFM signal components such as carrier frequency, linear FM ratio, amplitude and phase is of great importance for ISAR, cognitive radio, electronic countermeasures, star-earth communication, etc[7][8].

### 1.1.1 Linear Time-frequency Transform

At present, multicomponent time-varying signal parameter estimation mainly uses linear time-frequency transform and bilinear time-frequency transform and other methods. Among them, the linear time-frequency transform mainly includes the generalized Fourier transform GFT[9] (represented by the STFT, Gabor transform[10] and Stockwell transform[11]) and the wavelet transform WT with different basis functions.The generalized Fourier transform has clear physical significance in obtaining time-frequency domain information, and its inverse transform can realize the perfect reconstruction of signals under the COLA criterion[12]. Through post-processing techniques (e.g., Hough transform), the characteristic parameters of multicomponent time-frequency signals can be extracted approximately, and thus it has been widely used in the fields of speech signal processing and seismic signal processing[13]. However, the feature parameter extraction accuracy of its GFT algorithm relies heavily on the post-processing techniques, and there are problems such as low energy concentration; moreover, in the case of low signal-to-noise ratios, the blurring caused by the noise further increases the difficulty of accurately extracting the component feature parameters by the GFT method.

### 1.1.2 Bilinear Time-frequency Transform

Bilinear time-frequency transforms are mainly based on the WVD transform and its derived transforms, such as Pesudo WVD[14], Cohen's Class and the recently proposed LVD[], etc. The WVD transform has the theoretically highest time-frequency resolution due to the global characterization of its window function. However, in the process of solving the autocorrelation function, cross terms with different components are introduced. Especially in the case of more components, the amplitude of the cross terms in the time-frequency domain may even exceed the amplitude of the signal components themselves, thus seriously interfering with the extraction of the feature parameters. In order to solve this problem, many improved methods based on WVD, such as Pesudo WVD, Cohen's Class[15], LVD[16], and Gabor WVD based on the addition of windowing function, have been proposed. These methods better suppress the amplitude of the cross terms. However, like the linear time-frequency transform, the bilinear time-frequency transform cannot provide the parameters of the components directly. Therefore, it is still necessary to extract and fit the component information and parameters of the signal by corresponding post-processing transforms, such as the WVD-Hough transform, which reduces the accuracy of extracting the feature parameters to a certain extent. In addition, unlike the linear time-frequency transforms, the bilinear time-frequency transforms are not reversible, which results in the difficulty of processing the signals, such as filtering, noise reduction, and so on.

### 1.1.3 Mixed-frequency demodulation methods and Fractional-order Fourier transform methods

For the decomposition of multi-component linear FM signals, in addition to the generalized time-frequency transform methods mentioned above, there are also demodulation methods based on a priori knowledge and fractional-order Fourier transforms. The demodulation method based on a priori knowledge mixes the received time-delayed linear FM signal with the known signal, converts it into a sinusoidal signal and then estimates it, which is mainly used in the field of distance measurement. The fractional order Fourier transform FrFT, on the other hand, obtains the FM parameters by searching the fractional transform parameters in one dimension, and this method can improve the accuracy of obtaining the FM parameters by shortening the searching step, but at the cost of the rapid increase of the computational complexity. It is worth noting that the FrFT method can only obtain the FM parameters, but cannot directly obtain the information of other characteristic parameters such as the starting frequency, amplitude and initial phase.

## 1.2 Main achievements and innovations

Currently, almost all studies on multicomponent frequency modulation (LFM) signal decomposition algorithms are conducted under the condition of following Nyquist's theorem. However, with the development of technology and the expansion of application requirements, it is more and more difficult to achieve the sampling rate required by Nyquist due to the difficulty of sample acquisition and the limitation of hardware performance. For example, in some fields, the highest frequency of the signal can reach the terahertz level, which poses a great challenge to the traditional sampling and data processing and transmission. Therefore, it is particularly important to study the feature extraction algorithms for multicomponent FM signals under sub-Nyquist sampling conditions. At present, there are relatively limited studies on feature parameter extraction for sub-Nyquist sampling of multicomponent FM signals. Literature [] proposed a method to accurately estimate the feature parameters of its multicomponent linear FM signal components by converting the multicomponent linear FM signal into a multifrequency signal through hardware and using the compressed sensing (CS) method to extract the signal parameters. However, the method of sampling such signals under sub-Nyquist uniform sampling conditions and directly extracting all component parameters has not been reported. In this paper, a new multi-parameter feature parameter extraction algorithm is proposed for the first time, which successfully realizes the direct and accurate extraction of the feature parameters of multi-component LFM signals under very low sub-Nyquist sampling rate conditions.

In the problem of parameter extraction for multicomponent linear FM signals, the generalized eigenvalue method described in the literature [] is no longer applicable due to the extension of the studied parameters from individual frequency eigenvalues to eigenvalue pairs of the starting frequency and the FM parameter (i.e., chirp rate). For this reason, the method needs to be improved using special techniques in constructing and solving the equations to overcome this challenge. This is one of the main research and innovation points of this paper.

### 1.3 Organization

This paper is organized as follows: in Section 2, a brief review of the necessary prior knowledge such as derivative sampling theory and multi-parameter eigenvalue theory will be presented to ensure that the reader has a solid understanding of the foundational concepts that will be discussed subsequently. Subsequently, in Part III, the theoretical soundness of the algorithm proposed in this paper will be elaborated. This part will be developed in three segments. First, a system of equations will be constructed for multi-component linear FM signal feature parameter extraction, which is the core foundation of the algorithm. Second, a rigorous mathematical proof of the logic and validity of the algorithm will be carried out to verify its theoretical rationality and reliability. Finally, the signal decomposition strategy based on multi-parameter eigenvalue theory will be explored to further reveal the theoretical advantages of this paper's algorithm in dealing with complex problems. In the fourth part, it will turn to the performance study of this paper's method in practical applications. Specifically, the focus will be on the ability of the method to accurately extract the component feature parameters under sub-Nyquist sampling conditions, as well as its immunity to noise interference and enhancement path. In addition, the computational efficiency of the algorithm will be analyzed in depth and corresponding optimization strategies will be proposed, aiming to enhance the practicality and efficiency of the algorithm.

## 2 Prerequisite Knowledge

### 2.1 Derivative Sampling Theory

Derivative sampling is also essentially a multichannel sampling technique whose origins can be traced back to Shannon's pioneering work in 1949. Since then, the method has been further developed and refined by Fogel and Jagerman (1955, 1956). They proposed a specific implementation which requires simultaneous sampling of the signal and its first-order derivatives. This sampling strategy achieves a perfect reconstruction of the signal at half the sampling rate of the Nyquist sampling theorem.

According to the core principle of derivative sampling theory, if a signal is sufficiently smooth, there is enough information embedded in its higher-order derivatives to make it possible to reconstruct the original signal at a lower sampling rate. This approach cleverly exploits the intrinsic connection between a signal and its derivatives by obtaining samples of the signal's higher-order derivatives rather than the signal itself, thus achieving the goal of reducing the sampling rate.

Prior research evidence has shown that sampling based on the mth-order differentiation of a signal theoretically permits perfect signal reconstruction at $1/m$ of the Nyquist sampling rate. This finding provides an important theoretical basis and practical guidance for efficient, low-rate signal sampling and processing. Overall, derivative sampling, as an innovative sampling technique, opens up new possibilities for reducing the sampling rate while guaranteeing the reconstruction quality by digging deeper into the derivative properties of signals.

### 2.2 Multiparameter eigenvalue Theory

The multi-parameter eigenvalue problem can be regarded as an extension of the generalized eigenvalue algorithm, in which the object of study is the expansion of the scalar parameter corresponding to a single generalized eigenvalue into a tuple of eigenvalues of the form. Multi-parameter eigenvalue problem originates from the study of the parameter separation problem of Sturm-Liouville equation. Arscott, F. M. explicitly proposed the statement of two-parameter eigenvalue problem when he studied the differential problem in 1964 [1]; in the same year, Atkinson proposed the statement of multivariate spectral theory. The systematic study of multi-parameter eigenvalue problems is found in the related monograph of Atkinson, F.V [2].

Recently, Muhič and Plestenjak conducted an in-depth study on the solution methods of singular multiparameter eigenvalue problems, rectangular multiparameter eigenvalues, and the classification of solutions, and proposed a classification method for different types of dissimilar eigenvalues [3]. Dong B et al. studied the homogeneous solution of large-scale multiparameter eigenvalue problems and proposed a method that is universal and more effective for large order coefficient matrices [4]. In addition, Pons A, Gutschmidt S et al. delved into applications in non-aeroelastic modes by designing two iterative algorithms for the nonlinear multiparameter eigenvalue problem arising in the analysis of aeroelastic flutter, as well as a limiting method for simplifying the behavior of the system far away from the desired flutter point [5].

# 3 Theory of Truly sub-Nyquist multiparameter eigenvalue signal decomposition Method

## 3.1 Construction of equations for the sub-Nyquist multiparameter eigenvalue signal decomposition method

**Definition 3.1.** *Let a signal consisting of m linear FM signals $x_i(t)$ be $x(t)$, then it is of the following form:*

$$x(t)=\sum_{i=1}^{m} a_i e^{j(k_i t^2 + 2\pi f_i t + \varphi_{i0})} \tag{3.1}$$

where $a_i, k_i, f_i, \varphi_{i0}$ are the amplitude, FM index, starting frequency, magnitude and initial phase corresponding to the ith component, respectively, and j is an imaginary unit.

For simplicity let $\omega_i = 2\pi f_i$, then x(t) can be written in the following form:

$$x(t)=\sum_{i=1}^{m} a_i e^{j(k_i t^2 + \omega_i t + \varphi_{i0})} \tag{3.2}$$

Derivation of $x(t)$ gives its derivative $\dot{x}(t)$ as:

$$\begin{aligned} \dot{x}(t) &= \sum_{i=1}^{m} j(2k_i t + \omega_i) a_i e^{j(k_i t^2 + \omega_i t + \varphi_{i0})} \\ &= \sum_{i=1}^{m} j(2k_i t + \omega_i) x_i \\ &= \sum_{i=1}^{m} j 2k_i t x_i + \sum_{i=1}^{m} j \omega_i x_i. \end{aligned} \tag{3.3}$$

**Notation 3.1.** *Since the multi-parameter eigenvalue problem involved in this paper requires discrete matrices corresponding to different time intervals, the naming rules of the signals in this paper are specified as follows for ease of memorization and understanding:*

I. Continuous signals without explicitly declared time intervals are denoted by $x$ and its variants, e.g., a continuous interval signal is $x(t)$ with derivative $\dot{x}(t)$, the hankel matrix constructed from $x(t)$ discrete signals is $X$, and the hankel matrix constructed from $x(t)$ discrete signals is $\dot{X}$.

II. Continuous signals on explicitly declared time intervals are all represented using the lowercase font corresponding to the name of the corresponding interval and its variants, e.g., a continuous signal on the continuous interval $\mathcal{P}$ is $p(t)$, the derivative is $\dot{p}(t)$, the hankel matrix constructed from $p(t)$ discrete signals is $P$, and the hankel matrix constructed from $\dot{p}(t)$ discrete signals is $\dot{P}$.

III. Continuous time variables for which the time interval is not explicitly declared are denoted by $t$ and its variants, such as in equation (3.3), where t is used and the hankel matrix constructed discretely by $t$ is $T$.

Discretizing and constructing the hankel matrix for $x(t)$ over some continuous time interval as $X$, with the corresponding component matrix as $X_i$, discretizing and constructing the hankel matrix for $x(t)$ as $X$, with the corresponding component matrix as $X_i$, and discretizing and constructing the hankel matrix for $t$ at a sampling point as $T$, then $X$, $\dot{X}$ and $T$ have the following form.

$$X=\begin{pmatrix} x_1 & x_2 & \cdots & x_n \\ x_2 & x_3 & \cdots & x_{n+1} \\ \vdots & \vdots & \ddots & \vdots \\ x_n & x_{n+1} & \cdots & x_{2n-1} \end{pmatrix}, \dot{X}=\begin{pmatrix} \dot{x}_1 & \dot{x}_2 & \cdots & \dot{x}_n \\ \dot{x}_2 & \dot{x}_3 & \cdots & \dot{x}_{n+1} \\ \vdots & \vdots & \ddots & \vdots \\ \dot{x}_n & \dot{x}_{n+1} & \cdots & \dot{x}_{2n-1} \end{pmatrix}$$

$$T=\begin{pmatrix} t_1 & t_2 & \cdots & t_n \\ t_2 & t_3 & \cdots & t_{n+1} \\ \vdots & \vdots & \ddots & \vdots \\ t_n & t_{n+1} & \cdots & t_{2n-1} \end{pmatrix} \tag{3.4}$$

From equations (3.3) and (3.4), we have

$$\begin{aligned}\dot{X} &= \begin{pmatrix} \dot{x}_1 & \dot{x}_2 & \cdots & \dot{x}_n \\ \dot{x}_2 & \dot{x}_3 & \cdots & \dot{x}_{n+1} \\ \vdots & \vdots & \ddots & \vdots \\ \dot{x}_{n+1} & \dot{x}_{n+2} & \cdots & \dot{x}_{2n-1} \end{pmatrix} \\ &= j\begin{pmatrix} \sum_{i=1}^m (2k_i t_1+\omega_i)x_{i1} & \sum_{i=1}^m (2k_i t_2+\omega_i)x_{i2} & \cdots & \sum_{i=1}^m (2k_i t_n+\omega_i)x_{in} \\ \sum_{i=1}^m (2k_i t_2+\omega_i)x_{i2} & \sum_{i=1}^m (2k_i t_3+\omega_i)x_{i3} & \cdots & \sum_{i=1}^m (2k_i t_{n+1}+\omega_i)x_{i(n+1)} \\ \vdots & \vdots & \ddots & \vdots \\ \sum_{i=1}^m (2k_i t_n+\omega_i)x_{in} & \sum_{i=1}^m (2k_i t_{n+1}+\omega_i)x_{i(n+1)} & \cdots & \sum_{i=1}^m (2k_i t_{2n-1}+\omega_i)x_{i(2n-1)} \end{pmatrix} \\ &= j\sum_{i=1}^m 2k_i \begin{pmatrix} t_1 x_{i1} & t_2 x_{i2} & \cdots & t_n x_{in} \\ t_2 x_{i2} & t_3 x_{i3} & \cdots & t_{n+1} x_{i(n+1)} \\ \vdots & \vdots & \ddots & \vdots \\ t_n x_{in} & t_{n+1} x_{i(n+1)} & \cdots & t_{2n-1} x_{i(2n-1)} \end{pmatrix} + j\sum_{i=1}^m \omega_i \begin{pmatrix} x_{i1} & x_{i2} & \cdots & x_{in} \\ x_{i2} & x_{i3} & \cdots & x_{i(n+1)} \\ \vdots & \vdots & \ddots & \vdots \\ x_{in} & x_{i(n+1)} & \cdots & x_{i(2n-1)} \end{pmatrix} \\ &= j\sum_{i=1}^m 2k_i T\odot X_i + j\sum_{i=1}^m \omega_i X_i, \end{aligned} \tag{3.5}$$

where $\odot$ refers to the Harmard product.

For simplicity,

$$X_H = \sum_{i=1}^m X_{Hi} = \sum_{i=1}^m T\odot X_i. \tag{3.6}$$

Then it is obtained from equation (3.5) and equation (3.6):

$$\begin{aligned} \dot{X} &= j\sum_{i=1}^m 2k_i T\odot X_i + j\sum_{i=1}^m \omega_i X_i \\ &= j\sum_{i=1}^m 2k_i X_{Hi} + j\sum_{i=1}^m \omega_i X_i \end{aligned} \tag{3.7}$$

**Theorem 3.1.** *With finite precision, all the eigenparameters of $x(t)$ can be extracted when the sum of the ranks of all the component matrices $X_i$ of $X$ is less than or equal to the order of $X$. The modulation parameter $k_i$ of $x_i(t)$, the starting frequency $f_i$, is given by the following set of multiparameter generalized eigenvalue equations:*

$$\begin{cases} \dot{P}u = \lambda P_H u + \mu Pv \\ \dot{Q}v = \lambda Q_H v + \mu Qv \end{cases}, \tag{3.8}$$

where $\dot{P},\dot{Q}$ are the hankel matrices discretized and constructed for the signal of $x(t)$ on the intervals $\mathcal{P},\mathcal{Q}$. P,Q are the hankel matrices discretized and constructed for the signal of the multicomponent signal $x(t)$ on the intervals $\mathcal{P},\mathcal{Q}$ according to the naming notation 3.1 hankel matrices, $P_H, Q_H$ are the Harmard products of the time-discretized matrices $T_p, T_Q$ with P,Q on the corresponding intervals $\mathcal{P},\mathcal{Q}$ respectively. and $u,v$ are the corresponding eigenvectors.

Since the proof of Theorem 3.1 requires a proof of the premises of the theorem and of the lemma:

I. Proof of singularity or approximate singularity of $X_i$ with finite precision.

II. Proof of linear mapping relations between $P_H$ and $P$, $Q_H$ and $Q$.

Therefore, the proof of Theorem 3.1 will be placed after the completion of the above proof, followed by the completion of the premises and the lemma first.

## 3.2 Proof of the approximate singularity of the discrete composition of Hankel matrices for single-component linear FM signals with finite accuracy

Similar to the structure of multicomponent complex exponential signals, although the reconstructed Hankel matrix of a single-component linear FM signal is not a rank-one matrix due to its wider spectrum, if it can be proved that the rank of the matrix of a single-component linear FM signal is much lower than the order of the matrix in a particular case, the rank of the corresponding matrix of a multicomponent linear FM signal is still much lower than the order of the sum-signal Hankel matrix, which provides a theoretical possibility for the parameter's accurate extraction of separation provides a theoretical possibility.

A theoretical study on the non-full rank of single-component linear FM signal matrices with finite accuracy is proved as follows:

Since $x_i(t)$ is analytic everywhere in the complex plane, it is an integral function, so it is expanded based on the Taylor series, which is expanded as follows:

$$x_i(t) = \sum_{-\infty}^{+\infty} (a_n)_i t^n \tag{3.9}$$

Then the hankel matrix $X_i$ discretized and formed by $x_i(t)$ can be expressed in the following form:

$$X_i = \begin{pmatrix} \sum_{-\infty}^{+\infty}(a_n)_i t_{1i}^n & \sum_{-\infty}^{+\infty}(a_n)_i t_{2i}^n & \cdots & \sum_{-\infty}^{+\infty}(a_n)_i t_{ni}^n \\ \sum_{-\infty}^{+\infty}(a_n)_i t_{2i}^n & \sum_{-\infty}^{+\infty}(a_n)_i t_{3i}^n & \cdots & \Sigma(a_n)_i t_{(n+1)i}^n \\ \vdots & \vdots & \ddots & \vdots \\ \sum_{-\infty}^{+\infty}(a_n)_i t_{ni}^n & (a_n)_i t_{(n+1)i}^n & \cdots & \sum_{-\infty}^{+\infty}(a_n)_i t_{(2n+1)i}^n \end{pmatrix} \tag{3.10}$$

Considering the high starting frequency and small sampling interval, the higher terms of $\sum_{-\infty}^{+\infty}(a_n)_i t^n$ are geometrically decaying, when the higher terms are smaller than the lower limit of the system accuracy, so that X_i is approximately equal to a finite order polynomial, and according to [], the order of the hankel matrix composed of finite order polynomials is finite, so that Theorem 3.1 about the components holds in finite accuracy. Figure () demonstrates the variation of the ratio of rank to order corresponding to a single-component linear FM signal and its components for different sampling rates and different FM coefficients.

## 3.3 Theoretical study of linear mapping relationship between modulation parameter matrix and original signal matrix

Considering this sampling as equal interval continuous sampling then the time series is as follows:

$$t_g = g\Delta t \tag{3.11}$$

where $g=1,2,\cdots 2n-1$.

Then the component $X_H$ of $X_{iH}$ can be expressed in the following form, the

$$\begin{aligned}X_{iH} &= \Delta t \begin{pmatrix} 1x_{i1} & 2x_{i2} & \cdots & nx_{in} \\ 2x_{i2} & 3x_{i3} & \cdots & (n+1)x_{i(n+1)} \\ \vdots & \vdots & \ddots & \vdots \\ nx_{in} & (n+1)x_{i(n+1)} & \cdots & (2n-1)x_{i(2n-1)} \end{pmatrix} \\ &= \Delta t \begin{pmatrix} 1 & 0 & \cdots & 0 \\ 0 & 2 & \cdots & 0 \\ \vdots & \vdots & \ddots & \vdots \\ 0 & 0 & \cdots & n \end{pmatrix} \begin{pmatrix} x_{i1} & x_{i2} & \cdots & x_{in} \\ x_{i2} & x_{i3} & \cdots & x_{i(n+1)} \\ \vdots & \vdots & \ddots & \vdots \\ x_{in} & x_{i(n+1)} & \cdots & x_{i(2n-1)} \end{pmatrix} + \Delta t \begin{pmatrix} x_{i1} & x_{i2} & \cdots & x_{in} \\ x_{i2} & x_{i3} & \cdots & x_{i(n+1)} \\ \vdots & \vdots & \ddots & \vdots \\ x_{in} & x_{i(n+1)} & \cdots & x_{i(2n-1)} \end{pmatrix} \begin{pmatrix} 0 & 0 & \cdots & 0 \\ 0 & 1 & \cdots & 0 \\ \vdots & \vdots & \ddots & \vdots \\ 0 & 0 & \cdots & n-1 \end{pmatrix} \\ &= \Delta t D_{i1} \begin{pmatrix} x_{i1} & x_{i2} & \cdots & x_{in} \\ x_{i2} & x_{i3} & \cdots & x_{i(n+1)} \\ \vdots & \vdots & \ddots & \vdots \\ x_{in} & x_{i(n+1)} & \cdots & x_{i(2n-1)} \end{pmatrix} + \Delta t \begin{pmatrix} x_{i1} & x_{i2} & \cdots & x_{in} \\ x_{i2} & x_{i3} & \cdots & x_{i(n+1)} \\ \vdots & \vdots & \ddots & \vdots \\ x_{in} & x_{i(n+1)} & \cdots & x_{i(2n-1)} \end{pmatrix} D_{i2},\end{aligned} \quad (3.12)$$

Here:

$$D_{i1}=\begin{pmatrix} 1 & 0 & \cdots & 0 \\ 0 & 2 & \cdots & 0 \\ \vdots & \vdots & \ddots & \vdots \\ 0 & 0 & \cdots & n \end{pmatrix}, D_{i2}=\begin{pmatrix} 0 & 0 & \cdots & 0 \\ 0 & 1 & \cdots & 0 \\ \vdots & \vdots & \ddots & \vdots \\ 0 & 0 & \cdots & n-1 \end{pmatrix} \quad (3.13)$$

Consider the second half of Equation (3.12) and let the lth column of $X_i$ be $w_{il}$, where $l=1,2,\cdots,n$:

$$\begin{aligned}&\Delta t \begin{pmatrix} x_{i1} & x_{i2} & \cdots & x_{in} \\ x_{i2} & x_{i3} & \cdots & x_{i(n+1)} \\ \vdots & \vdots & \ddots & \vdots \\ x_{in} & x_{i(n+1)} & \cdots & x_{i(2n-1)} \end{pmatrix} \begin{pmatrix} 0 & 0 & \cdots & 0 \\ 0 & 1 & \cdots & 0 \\ \vdots & \vdots & \ddots & \vdots \\ 0 & 0 & \cdots & n-1 \end{pmatrix} \\ &= \Delta t [0w_{i1}, 1w_{i2}, \cdots, (n-1)w_{in}]\end{aligned} \quad (3.14)$$

Next it will be shown that for a matrix $X_i, D_{i2}$ there exists a matrix $X_{Ri}$, satisfying the following relation:

$$X_{Ri}X_i = X_i D_{i2} \quad (3.15)$$

Consider that the above three satisfy the following relationship:

$$\begin{aligned} X_{Ri}X_i &= [X_{Ri}w_{i1}, X_{Ri}w_{i2}, \cdots, X_{Ri}w_{in}] \\ X_i D_{i2} &= [0w_{i1}, 1w_{i2}, \cdots, (n-1)w_{in}] \end{aligned} \quad (3.16)$$

Equalizing the right-hand side counterpart of Eq.(3.15) yields:

$$[X_{Ri}w_{i1}, X_{Ri}w_{i2}, \cdots, X_{Ri}w_{in}] = [0w_{i1}, 1w_{i2}, \cdots, (n-1)w_{in}] \quad (3.17)$$

Then from equation (3.17), we can get:

$$X_{Ri}w_{i1}=0w_{i1}, X_{Ri}w_{i2}=1w_{i2}, \cdots, X_{Ri}w_{in}=(n-1)w_{in} \quad (3.18)$$

From equation (3.18) it follows that i.e. $X_{Ri}$ is a matrix with l-1 as eigenvalues and $w_{i2}$ as the corresponding eigenvectors. the existence of $X_{Ri}$ is proved.

And:

$$X_{Mi}=D_{i1}+X_{Ri} \quad (3.19)$$

Then, from the definition of $X_{Hi}$ and Equations (3.12) and (3.19), we have

$$X_H=\sum_{i=1}^{m} X_{Hi}=\sum_{i=1}^{m} X_{Mi}X_i \quad (3.20)$$

## 3.4 Theorem of parameter extraction of multi-component linear frequency modulation signal components

In this section, the proof of Theorem 3.1 will be completed in Sections 3.2 and 3.3 based on the conclusion that, with limited precision, the discrete composition of a single-component linear frequency modulation signal Hankel matrix is approximately singular and the linear mapping relationship between the modulation parameter matrix and the original signal matrix.

First, the diagonalization of the weight matrix $X_i$ is performed. Considering that the matrix constructed in this paper is a complex Hankel matrix, which is a complex symmetric matrix but not an Hermitian matrix, the diagonalization method for Hermitian matrices cannot be used. Fortunately, we can still use the Takagi Factorization[] method, also known as the SSVD decomposition[] method, to diagonalize $X_i$:

$$X_i = V_i^T S_i V_i \tag{3.21}$$

Where $V_i$ is the right singular vector matrix of $X_i$ is a You matrix, $S_i$ is the diagonal matrix of $X_i$ after SSVD decomposition, it should be noted that here the left hand side is no longer complex conjugate but ordinary transpose.

Considering that the component matrix $\dot{X}_i$ of $X$ is a linear combination of the component $X_{Hi}$ of $X_H$ with respect to the FM exponent $k_i$, and the component matrix $X_i$ of $X$ with respect to the starting frequency $\omega_i$, the right-hand side of Eq. (3.7) can be expressed in the following form by Eq.(3.20), Eq.(3.21):

$$[(k_1 X_{M1}+\omega_1 I_1)V_1, (k_2 X_{M2}+\omega_2 I_2)V_2, \cdots, (k_m X_{Mm}+\omega_m I_m)V_m]^T \begin{pmatrix} S_1 & 0 & \cdots & 0 \\ 0 & S_2 & \cdots & 0 \\ \vdots & \vdots & \ddots & \vdots \\ 0 & 0 & \cdots & S_m \end{pmatrix} [V_1, V_2, \cdots, V_m] \tag{3.22}$$

Similarly, $\lambda X_H + \mu X$ can be written in the following form:

$$[(\lambda X_{M1}+\mu I_1)V_1, (\lambda X_{M2}+\mu I_2)V_2, \cdots, (\lambda X_{Mm}+\mu I_m)V_m]^T \begin{pmatrix} S_1 & 0 & \cdots & 0 \\ 0 & S_2 & \cdots & 0 \\ \vdots & \vdots & \ddots & \vdots \\ 0 & 0 & \cdots & S_m \end{pmatrix} [V_1, V_2, \cdots, V_m] \tag{3.23}$$

Then from equations (3.22) and (3.23):

$$\begin{aligned}&(\dot{X} - \lambda X_H - \mu X)u \\ &= \begin{pmatrix} (k_1 X_{M1}+\omega_1 I_1-\lambda X_{M1}-\mu I_1)V_1 \\ \vdots \\ (k_i X_{Mi}+\omega_i I_i-\lambda X_{Mi}-\mu I_i)V_i \\ \vdots \\ (k_m X_{Mm}+\omega_m I_m-\lambda X_{Mm}-\mu I_m)V_m \end{pmatrix}^T \begin{pmatrix} S_1 & \cdots & 0 & \cdots & 0 \\ \vdots & \ddots & \vdots & \cdots & \vdots \\ 0 & \cdots & S_i & \cdots & 0 \\ \vdots & \vdots & \vdots & \ddots & \vdots \\ 0 & \cdots & 0 & \cdots & S_m \end{pmatrix} [V_1, \cdots, V_i, \cdots V_m]u\end{aligned} \tag{3.24}$$

Considering the correspondence, and the fact that Eq. (3.24) holds at any point in time, the corresponding $X_i$ is arbitrary, which in turn leads to the arbitrariness of $X_i$, so that when and only when $\lambda_i = k_i, \mu_i = \omega_i$时, $\dot{X} - \lambda X_H - \mu X$ corresponds to a non-full rank, i.e., its zero space exists, there exists a non-zero vector $u$, such that Eq. (3.25) holds:

$$(\dot{X} - \lambda X_H - \mu X)u = 0 \tag{3.25}$$

And combining with the classification theory of multi-parameter eigenvalues in the literature [][][], it can be seen that $\lambda_i, \mu_i$ are finite regular multi-parameter eigenvalues that satisfy the above equation. Since $P, Q, P_H, Q_H, \dot{P}, \dot{Q}$ are subsets of $X, X_H, \dot{X}$, the system of equations (3.8) in Theorem 3.1 also satisfies the property of Equation (3.25), and Theorem 3.1 is proved.

Regarding the solution of the system of equations (3.8), you can refer to the two-parameter eigenvalue solution method in the Preparatory Knowledge 2.2 as follows, based on the theory of multi-parameter eigenvalue, first construct the following equations to solve for $\Delta_0, \Delta_1, \Delta_2$:

$$\begin{cases} \Delta_0 = P_H \otimes Q - P \otimes Q_H \\ \Delta_1 = \dot{P} \otimes Q - P \otimes \dot{Q} \\ \Delta_2 = P_H \otimes \dot{P} - \dot{Q} \otimes Q_H \end{cases} \quad (3.26)$$

Here $\otimes$ in the above equation denotes the kroneker product.

Then, based on the above equation (3.26), the system of equations (3.8) can be transformed into the following system of generalized characteristic equations:

$$\begin{cases} \Delta_1(u \otimes v) - \lambda(u \otimes v) = 0 \\ \Delta_2(u \otimes v) - \mu(u \otimes v) = 0 \end{cases} \quad (3.27)$$

Then the value of $\lambda, \mu$ can be found from equation (3.27) based on the theory of generalized characteristic equations, and based on Theorem 3.1, the value of $\lambda, \mu$ is the FM exponent $k_i$ corresponding to the component $x_i(t)$ and $2\pi$ times $\omega_i$ of the starting frequency $f_i$.

## 3.5 Theory of accurate extraction of component amplitudes and initial phases of multicomponent linear FM signals

**Definition 3.2.** *Let the signal l(t) be*

$$l_i(t) = e^{j(k_i t^2 + 2\pi f_i t)}. \quad (3.28)$$

where $k_i$ and $f_i$ denote the corresponding frequency-modulation index and start frequency, respectively. The Hankel matrix constructed from discretized $l_i(t)$ is denoted as $L_i$, where

$$L_i = \begin{pmatrix} l_{i_1} & l_{i_2} & \cdots & l_{i_n} \\ l_{i_2} & l_{i_3} & \cdots & l_{i_{n+1}} \\ \vdots & \vdots & \ddots & \vdots \\ l_{i_n} & l_{i_{n+1}} & \cdots & l_{i_{2n-1}} \end{pmatrix}$$

**Corollary 3.1.** *Construct a generalized eigenvalue equation in the form of*

$$X_i v = \lambda_{i_{Ap}} L_i v \quad (3.29)$$

Once the start frequency and frequency-modulation index of the chirp signal are determined via Proposition 2, the amplitude and initial phase can be calculated.

**Proof.** From the definitions of $Y_i$ and $L_i$, it can be concluded that $\square$

$$y_i = a_i e^{j\varphi_{i0}} l_i \quad (3.30)$$

$$\begin{aligned}(X - \beta L_i) v \\ = \begin{pmatrix} V_1 \\ \vdots \\ V_i \\ \vdots \\ V_m \end{pmatrix}^T \begin{pmatrix} (\beta - a_i e^{j\varphi_0}) S_1 & \cdots & 0 & \cdots & 0 \\ \vdots & \ddots & \vdots & \cdots & \vdots \\ 0 & \cdots & (\beta - a_i e^{j\varphi_0}) S_i & \cdots & 0 \\ \vdots & \vdots & \vdots & \ddots & \vdots \\ 0 & \cdots & 0 & \cdots & (\beta - a_i e^{j\varphi_0}) S_m \end{pmatrix} [V_1, \cdots, V_i, \cdots V_m] v \end{aligned} \quad (3.31)$$

So that is when $\beta = a_i e^{j\varphi_{i0}}$ there exists a non-zero vector that satisfies $(X - \beta X_i)v$, According to generalized eigenvalue theory, $a_i e^{j\varphi_{i0}}$ is the generalized eigenvalue of $(X - \beta L_i)v = 0$.